%% file: main.tex
\documentclass[copyright,creativecommons]{eptcs}
\providecommand{\event}{ICE 2019} 
\usepackage{breakurl}             

\usepackage{booktabs}   
\usepackage{subcaption} 

\usepackage{amsmath}
\usepackage{amsthm}
\usepackage{amssymb}

\usepackage[british]{babel}

\usepackage{semantic}
\usepackage{stmaryrd}
\usepackage{color}
\usepackage{listings}
\usepackage{mathpartir}

\usepackage{pifont}

\usepackage{tcolorbox}
\usepackage{tabularx}
\usepackage{array}
\usepackage{colortbl}
\tcbuselibrary{skins}

\usepackage{longtable}

\usepackage{hyphenat}
\usepackage{hyperref}
\usepackage{flushend}

\newtheorem{Remark}{Remark}[section]

\input{definitions.tex}

\title{Towards Gradually Typed Capabilities in the Pi-Calculus}
\author{Matteo Cimini
\institute{University of Massachusetts Lowell\\ Lowell, MA, USA}
\email{matteo\_cimini@uml.edu}
}
\def\titlerunning{Towards Gradually Typed Capabilities  in the Pi-Calculus}
\def\authorrunning{Matteo Cimini}

\begin{document}

\maketitle

\begin{abstract}
Gradual typing is an approach to integrating static and dynamic typing within the same language, and puts the programmer in control of which regions of code are type checked at compile-time and which are type checked at run-time. 

In this paper, we focus on the $\pi$-calculus equipped with types for the modeling of input-output capabilities of channels.
We present our preliminary work towards a gradually typed version of this calculus. 
We present a type system, a cast insertion procedure that automatically inserts run-time checks, and an operational semantics of a $\pi$-calculus that handles casts on channels. 
Although we do not claim any theoretical results on our formulations, we demonstrate our calculus with an example and discuss our future plans. 
\end{abstract}

\section{Introduction}
 
This paper presents preliminary work on integrating dynamically typed features into a typed $\pi$-calculus. 
Pierce and Sangiorgi have defined a typed $\pi$-calculus in which channels can be assigned types that express input and output capabilities \cite{tp}. Channels can be declared to be input-only, output-only, or that can be used for both input and output operations. 
As a consequence, this type system can detect unintended or malicious channel misuses at compile-time. 

The static typing nature of the type system prevents the modeling of scenarios in which the input-output discipline of a channel is discovered at run-time. 
In these scenarios, such a discipline must be enforced with run-time type checks in the style of dynamic typing. 


Gradual typing provides a natural solution to do just that, as it advocates the integration of dynamic typing and static typing \cite{Siek:2006bh,Tobin-Hochstadt:2006}. In gradual typing the programmer is in control of which regions of code are type checked at compile-time and which are given a pass, with the promise to perform checks at run-time when actual values are available. 

%

In this paper, we present our preliminary work to equip Pierce's and Sangiorgi's typed $\pi$-calculus with gradual typing. Our contributions are: 

\begin{itemize}
\item We present a formulation of a gradual type system for capabilities in $\pi$-calculus. This type system allows for channels to be declared dynamically typed, that is, their capabilities will be only known at run-time. Accordingly, the type system is permissive when dynamically typed channels occur, but still rejects processes if strong inconsistencies are found at compile-time. 

\item We offer a cast insertion procedure: A run-time check is inserted at any spot in which the gradual type checker has been optimistic and gave a pass. We make use of casts on channels to perform run-time checks. 

\item We present a formulation of a $\pi$-calculus with cast channels. This is an operational semantics that extends that of the $\pi$-calculus to handle casts on channels and detect when they fail or succeed. 
\end{itemize}

In this paper we do not claim any theoretical results on our formulations, hence the ``\emph{Towards}'' part of the title. We offer this preliminary work, demonstrate it with an example, and  discuss our future plans.  

The following sections are organized as follows. Section \ref{background} reviews the typed $\pi$-calculus of Pierce and Sangiorgi and motivates the benefits of adding gradual typing. Section \ref{gt} shows our formalisms for gradually typed capabilities in $\pi$-calculus. 
Section \ref{example} applies our formalisms to an example. 
Section \ref{properties} discusses the properties that we plan to establish in the future. 
Section \ref{related} discusses related work, and Section \ref{conclusions} concludes the paper.


\section{Background}
\label{background}
We review the type system of Pierce's and Sangiorgi's typed $\pi$-calculus with the example that the authors offered in \cite{tp}. The scenario is that of a printer server $P$ that interacts with multiple clients $C_1$, \ldots, $C_n$. The printer provides a channel $p$ over which clients can communicate their job requests. The printing system can be described with the process $\nu p. (P \mid C_1 \mid C_2 \mid \ldots \mid C_n)$. Suppose a client $C_1 \equiv \overline{p}\langle j_1\rangle.\overline{p}\langle j_2\rangle$ requests jobs $j_1$ and $j_2$ to be printed. A malicious client $C_2 \equiv {p}(j).C_2$ could very well interject the job requests of $C_1$, and pretend to be the printer. 
In the $\pi$-calculus with capabilities this scenario can be prevented by enforcing that clients used $p$ only in output operations. More generally, a channel can be given three kinds of capability types\footnote{We use a slightly different notation than that of \cite{tp}.}: $\out \cdot (T_1, \ldots, T_n)$, which means that the channel can be used output-only for $n$ channels of types $T_1, \ldots, T_n$ (these types are capabilities themselves), $\inp \cdot (T_1, \ldots, T_n)$, which means that the channel can be used input-only for $n$ arguments of types $T_1, \ldots, T_n$, and $\either \cdot (T_1, \ldots, T_n)$, which means that the channel can be used both for input and output operations. 
If we were to use the type system to the printing example, we would type check the clients separately, and would do so to impose the restriction that $p$ must be of type $\out \cdot (T)$, where $T$ is the type of the print request (irrelevant here). Thanks to this typing discipline the malicious client $C_2$ would be rejected. The printer system has good guarantees just by knowing that clients type check successfully, without inspecting the code of the clients. 
 

\paragraph{Decisions at Run-time}
Let us consider the following example. After submitting a tax return, a tax payer either receives a refund or must send a payment to a revenue agency. For the sake of the example, suppose that the revenue agency operates with limited resources and can only share one channel ($b$ below) with the tax payer. This channel is then used to either receive or send a payment. 
The tax payer is left free to write the application that interacts with the revenue agency. Because of this, we certainly should impose a suitable discipline on channels, or we may leave ourselves open to malicious scenarios akin to those of the printer example. Therefore, we shall use the $\pi$-calculus with capability types. 

The revenue agency $A$ can be modeled as 
\begin{flalign}
A  \equiv\app &\nu(x:\out \cdot T).\overline{r}\langle x\rangle.\overline{x}\langle \$100\rangle \\\nonumber
&+\\\nonumber
&\nu(x:\inp \cdot T).\overline{r}\langle x\rangle.x(sum)
\end{flalign}
 
The channel $r$ sends the newly created channel $x$ to the client. The type $T$ is the type of the channel representing money (irrelevant here). 
\begin{Remark}\label{remark}
To simplify our example, we assume that the choice operation $+$ picks the correct branch (the branch for sending a tax refund or the branch for waiting for a payment). 
\end{Remark}

In our situation, which branch of the $+$ operation will fire will be known at run-time. A client $C$ can be modeled as follows. 
\begin{equation}
C \equiv t(b: \textit{what type?}). ~ (\overline{b}\langle \$100\rangle ~+~ b(sum))\\
\end{equation}

After receiving the channel $b$, the tax payer client is ready to either send or receive payments, depending on what action the revenue agency makes available. 

\emph{What type should we assign to the channel $b$?} Were we to declare $b$ to be input-only, the process $C$ would be rejected because it uses $b$ in an output operation (in $\overline{b}\langle \$100\rangle$). Were we to declare $b$ to be output-only, $C$ would be similarly rejected. If we want to use the type system with capabilities and have a chance to pass the type checker we have no choice but to declare channel $b$ as $\either$. However, this has the effect to simply leave the tax payer client free to use $b$ with no restrictions. The client application could then exploit unintended mistakes of the revenue agency process. Consider for example the following client.
\begin{equation}
C \equiv t(b:\either \cdot T). ~ (\overline{b}\langle \$100\rangle\HI{$.b(sum)$} ~+~ b(sum))\\
\end{equation}

This version of the client is well-typed in the typed $\pi$-calculus of Pierce and Sangiorgi. 
Furthermore, it does, in all appearances, what is expected: It does pay the whole sum when asked to, and it does receive the whole sum when asked to. However, when the client is mandated to send payments, it silently also tries to sneak in a reading operation on channel $b$, on the off chance that the revenue agency reuses the same channel by mistake and some other tax payer would send money over $b$. 

Allowing $\either$ is not an option, and neither it is the use of any of the other types. 

\paragraph{Integrating Gradual Typing}

The problem with the example above is that the type of channel $b$ is not known at compile-time. On the contrary, it can only be discovered at run-time. We therefore integrate Pierce's and Sangiorgi's typed $\pi$-calculus with dynamic typing features, where type checking is postponed at run-time. We do so by adopting the gradual typing style, an approach for integrating static and dynamic typing within the same language \cite{Siek:2006bh}. 

We introduce the possibility to assign the dynamic type $\dyn$ (\cite{Harper}) to channels. This means that the capability of the channel is not known at compile time. When the channel is used at run-time it will find itself instantiated with a well-determined channel, and the moment this channel is used we verify that its capabilities allow for the operation required. 

In our calculus, the client application can be modeled as follows. 
\begin{equation}
C \equiv r(b:\HI{$\dyn$} ). ~ (\overline{b}\langle \$100\rangle ~+~ b(sum))\\
\end{equation}

When the type checker encounters the sub-expressions $\overline{b}\langle \$100\rangle$ and $b(sum)$ it essentially gives them a pass, as they cannot be resolved at compile time. 
However, those checks must be performed at run-time. To perform such checks we follow the standard approach in gradual typing, thus we have a compilation step that inserts run-time checks. 
A popular mechanism for performing these checks is through the means of casts \cite{SiekTW15,SiekW10}. This procedure is typically called \emph{cast insertion} in gradual typing. After cast insertion, the client becomes
\begin{gather}
(r(b:\dyn). ~ (\overline{(\HI{$b:\dyn \Rightarrow \out \cdot T$})}\langle \$100\rangle ~+~ (\HI{$b:\dyn \Rightarrow \inp \cdot T$})(sum))
\end{gather}

\emph{Cast channels} are highlighted: Casts wrap around channels and have the form $a : T_1 \Rightarrow T_2$, which means that the channel $a$ is of type $T_1$ and is cast to the type $T_2$. 
Cast channels are used in lieu of ordinary channels as the subject of input and output operations, and they can also be communicated in output. 
Above, $b:\dyn \Rightarrow \out \cdot T$ means that at compile-time we only knew that $b$ was of dynamic type but at run-time it needs to be a channel with output capabilities. $b:\dyn \Rightarrow \inp \cdot T$ is analogous.  

When using Pierce's and Sangiorgi's calculus we frequently type check a part of the process under some capability assignments, and other parts with another. For example, we type check the clients separately. It makes little sense for the printer in the printer example to be type checked with the same assignments as clients, as types must be dual, at least. 
We inherit the same workflow. However, we need some extra care because the server needs to tell clients that some specific channels must be used with the opposite operation than that which the server performs. The server needs to advertise a reversed type for a channel. To enable this scenario, 
we augment the $\pi$-calculus with a tagged output $\stackrel{\rule{0.2cm}{0.4pt}_{{\mathcal{R}}}\hspace{-0.2cm}}{{r}}\langle x\rangle$. 

The agency process is then the following. 

\begin{flalign}
A  \equiv\app &\nu(x:\out \cdot T).\HI{$\stackrel{\rule{0.2cm}{0.4pt}_{{\mathcal{R}}}\hspace{-0.2cm}}{{r}}\langle x\rangle$}.\overline{x}\langle \$100\rangle \\\nonumber
&+\\\nonumber
&\nu(x:\inp \cdot T).\HI{$\stackrel{\rule{0.2cm}{0.4pt}_{{\mathcal{R}}}\hspace{-0.2cm}}{{r}}\langle x\rangle$}.x(sum)
\end{flalign}

To make an example, when the first branch of the choice operator fires the channel $x$ is declared output-only. 
The output $\stackrel{\rule{0.2cm}{0.4pt}_{{\mathcal{R}}}\hspace{-0.2cm}}{{r}}\langle x\rangle$ sends the channel $x$ but also informs the receiver that $x$ must be used as input-only. (The second branch follows analogous lines.) 
We achieve this scenario by having the compilation step inserting a cast to that effect. Intuitively, the agency  will have the following casts
\begin{flalign}
 &\nu(x:\out \cdot T).\overline{r}\langle \HI{$(x : \inp \cdot T \Rightarrow \dyn)$}\rangle.\overline{x}\langle \$100\rangle \\\nonumber
&+\\\nonumber
&\nu(x:\inp \cdot T).\overline{r}\langle \HI{$(x : \out \cdot T \Rightarrow \dyn)$}\rangle.x(sum)
\end{flalign}

Besides the casts, it is to notice that the special reverse output, precious in driving the cast insertion, has been compiled away into an ordinary output. 

When we run $(A \mid C)$ with the casts as in (5) and (7), we need to do so in an operational semantics that handles cast channels. We formulate such semantics in Section \ref{gt}. This formulation is capable of checking when casts succeed or fail at run-time. To make an example, let us suppose that the second branch of (7) occurs, i.e. the tax payer owes a payment. In one step the client would receive a cast channel in input and would become:  
(As a notational convenience, we collapse sequences of casts such as $(x:\out \cdot T \Rightarrow \dyn) : \dyn \Rightarrow \out \cdot T$ with the notation $x: \out \cdot T \Rightarrow \dyn \Rightarrow \out \cdot T$.

\[\overline{(\HI{$x: \out \cdot T \Rightarrow \dyn \Rightarrow \out \cdot T$})}\langle \$100\rangle ~+~ (\HI{$x:\out \cdot T \Rightarrow\dyn \Rightarrow \inp \cdot T$})(sum)
\]

Notice that the whole channel, with the cast around, has been passed. 
Now this process must interact with the remaining part of the agency, which is $x(sum)$.
However, the subject of the output $x: \out \cdot T \Rightarrow \dyn \Rightarrow \out \cdot T$ is not a bare channel. Therefore our operational semantics resolves these casts first, and checks that the channel buried inside the casts actually is of type $\out \cdot T$, which is requested at the end target of the cast. This is the case, and our operational semantics simply removes casts, which leads the client to take a step to $\overline{x}\langle \$100\rangle$. 
At this point, the ordinary communication rule of the $\pi$-calculus can take place. 
If, by mistake, another client $\overline{x}\langle \$5000\rangle$ were around, $C$ could not take its payment. Indeed, if the branch $(x:\out \cdot T \Rightarrow\dyn \Rightarrow \inp \cdot T)(sum)$ were  selected for a communication, our operational semantics first would detect that an output-only channel is being used for an input operation, and would trigger a cast error at run-time. 

The ability of handling capabilities that are unknown at compile-time, established at run-time, and protected during execution, is not possible in the $\pi$-calculus of Pierce and Sangiorgi. We have therefore enhanced that calculus with gradual typing in the next section. 

\begin{figure}[tbp]
\small
\begin{syntax}
  &&&\\
  \text{\sf Types} & T,S & ::= &  I \cdot (T_1, \ldots, T_n) \mid \HI{$\dyn$}\\
  \text{\sf Capabilities} & I & ::= &  \inp \mid \out \\
   \text{\sf Processes} & P & ::= & \nilProcess \mid ~ a(a_1:T_1, \ldots, a_n:T_n).P ~\mid~ \overline{a} \langle a_1, \ldots, a_n\rangle.P ~\mid \HI{$\stackrel{\rule{0.2cm}{0.4pt}_{{\mathcal{R}}}\hspace{-0.2cm}}{{a}}\langle a_1, \ldots, a_n\rangle.P$} \\
   &&& \mid~ P \mid P \mid~ P + P ~\mid~ (\nu a:T).P ~\mid~ !P \\
   \text{\sf Type Environment} & \Gamma & ::= & \emptyset \mid \Gamma, a:T
   \end{syntax}\\
   
\textsf{Type System}  \hfill  \fbox{$\Gamma \vdash P : \ok$}

\begin{gather*}
	{ \Gamma \typeOf \app \nilProcess : \ok}
\\[2ex]
\inference
	{
	\Gamma \typeOf \app P_1 : \ok &
	\Gamma \typeOf \app P_2 : \ok 	
	} 
	{ \Gamma \typeOf \app P_1+ P_2 : \ok}
\qquad 
\inference
	{
	\Gamma \typeOf \app P_1 : \ok &
	\Gamma \typeOf \app P_2 : \ok 	
	} 
	{ \Gamma \typeOf \app P_1\mid P_2 : \ok}
\\[2ex]
\ninference{t-res}
{\Gamma, a:T \typeOf \app P : \ok 
}
{\Gamma \typeOf  (\nu a:T).P : \ok }
\quad\,\,
\inference
	{
	\Gamma \typeOf \app P : \ok 	} 
	{ \Gamma \typeOf \app !P : \ok}
\\[2ex]
\ninference{t-in}
	{\HI{$\Gamma(a)  \sim \inp \cdot (T_1 \cdot  \ldots \cdot T_n)$} \qquad
	\Gamma, a_1:T_1, \ldots, a_n:T_n \typeOf \app P : \ok  	} 
	{ \Gamma \typeOf \app a(a_1:T_1, \ldots, a_n:T_n).P : \ok}
\\[2ex]
\ninference{t-out}
	{\HI{$\Gamma(a)  \sim \out \cdot (\Gamma(a_1)\cdot \ldots \cdot  \Gamma(a_n))$} \qquad
	\Gamma \typeOf \app P : \ok  	} 
	{ \Gamma \typeOf \app \{\overline{a} \app \key{or}\app \stackrel{\rule{0.2cm}{0.4pt}_{{\mathcal{R}}}\hspace{-0.2cm}}{{a}}\app\app\} \langle a_1, \ldots, a_n\rangle.P : \ok}
\end{gather*}
\textsf{Type Consistency}  \hfill  \fbox{$T \sim T$}
\begin{gather*}
T \sim \dyn \qquad \dyn \sim T 
\\[2ex]
\inference
{T_1 \sim S_1 & \cdots & T_n \sim S_n}
{\inp \cdot (T_1, \ldots, T_n) \sim \inp \cdot (S_1, \ldots, S_n) }
\\[2ex]
\inference
{T_1 \sim S_1 & \cdots & T_n \sim S_n}
{\out \cdot (T_1, \ldots, T_n) \sim \out \cdot (S_1, \ldots, S_n) }
\end{gather*}
\caption{Syntax and Type System of Gradually Typed Capabilities}
\label{fig:gt}
\end{figure}

\section{Gradually Typed Capabilities in the Pi-calculus}
\label{gt}

\subsection{Gradual Type System}
In this section we present our formalisms for a $\pi$-calculus with gradually typed capabilities. Fig. \ref{fig:gt} shows the syntax and the type system of our calculus, $\pi^{\dyn}$. The figure highlights the relevant parts of the system. 
The syntax is based on that of \cite{tp} (though we omit recursive types, as discussed in Section \ref{related}). 
The additions are the type $\dyn$ and the reverse output described in the previous section. 
The type system has the form $\Gamma \vdash P : \ok$, where $\Gamma$ contains associations from channels to their capabilities. This typing judgement means that under the assignments of $\Gamma$ the process $P$ is well-typed. 
Programmers type check their process $P$ by first providing $\Gamma$ with the capability associations for free channels. 
Further associations may be added during type checking when the type system traverses new bindings in restrictions $\nu$ and input operations.

Most of the design of Fig. \ref{fig:gt} mirrors the formulation of Pierce and Sangiorgi \cite{tp}, and much credits go to that formulation. 
One of the main differences is that our typing rules make use of type consistency $\sim$ (\cite{Siek:2006bh,AndersonD03}) in rules \textsc{(t-in)} and \textsc{(t-out)}, whereas the type system in \cite{tp} makes use of subtyping. 
Type consistency $T_1 \sim T_2$ holds when the two types are recursively the same type modulo the fact that one of them may have $\dyn$ instead of a concrete type. This is key to model the fact that the type system is liberal at the encounter of $\dyn$. 
We have that $\dyn \sim \out \cdot (T_1, \ldots, T_n)$, as well as $\out \cdot (T_1, \ldots, T_n) \sim \out \cdot (T_1, \ldots, T_n)$. However, the relation is not too liberal, and we have $\inp \cdot (T_1, \ldots, T_n) \not\sim \out \cdot (S_1, \ldots, S_n)$.  
In contrast to subtyping, $\sim$ is symmetric, and is not transitive. 

\subsection{Cast Insertion Procedure}

The gradual type system takes a permissive view at the encounter of $\dyn$. 
However, the checks that are not possible at compile-time must take place at run-time. 
For this reason we compile the program and insert casts. 
Fig. \ref{fig:ci} shows the cast insertion procedure, and highlights the casts that are inserted. This procedure is formulated with a judgement of the form $\Gamma \vdash P \leadsto P' : \ok$, which means that under the assignments declared in $\Gamma$, $P$ is well-typed and compiles into $P'$. $P'$ is essentially $P$ but contains casts around channels. A cast $a : T_1 \Rightarrow T_2$ means that the channel $a$ is of type $T_1$ and is cast to the type $T_2$. 
 Notice that $P'$ is not a process of $\pi^{\dyn}$, as $\pi^{\dyn}$ does not have a cast operator. $P'$ is a process of the calculus $\pi_c^{\dyn}$, which we define in the next section and which does have casts. 
In a nutshell, $\pi^{\dyn}$ acts as the surface language that programmers use. Ideally, its counterpart $\pi_c^{\dyn}$ with casts is not visible to programmers, though it is the end calculus that executes programs. Essentially, programmers do not handle casts explicitly.  

In Fig. \ref{fig:ci} casts are placed when two types are compared by $\sim$ rather than plain equality.  Furthermore, the reverse output generates a cast around for the output channel to label it as sending channels with the reverse type. As we shall see, from that point there is a reduction step that moves these casts to the arguments, so that the arguments are set to advertise their reverse type to the receiver. In Fig \ref{fig:ci}, the operation $\Gamma(a)^{-1}$ provides the reverse type of a capability. 
Once the cast is generated, the reverse output $\stackrel{\rule{0.2cm}{0.4pt}_{{\mathcal{R}}}\hspace{-0.2cm}}{{a}}\app$ is compiled away and becomes an ordinary output $\overline{a}$. 
It is worth pointing out that implementations can detect when casts are trivial, i.e. $c:T\Rightarrow T$, and omit generating them. 

\begin{figure}[tbp]
\small
\textsf{Cast Insertion}  \hfill  \fbox{$\Gamma \vdash P \leadsto P : \ok$}
\begin{gather*}
	{ \Gamma \typeOf \app \nilProcess \leadsto \nilProcess : \ok}
\\[2ex]
\inference
	{
	\Gamma \typeOf \app P_1 \leadsto P_1' : \ok &
	\Gamma \typeOf \app P_2 \leadsto P_2' : \ok 	
	} 
	{ \Gamma \typeOf \app P_1\mid P_2 :\leadsto P_1'\mid P_2'  \ok}
\qquad
\inference
	{
	\Gamma \typeOf \app P_1 \leadsto P_1' : \ok &
	\Gamma \typeOf \app P_2 \leadsto P_2' : \ok 	
	} 
	{ \Gamma \typeOf \app P_1+ P_2 :\leadsto P_1'+ P_2'  \ok}
\\[2ex]
\inference
{\Gamma, a:T \typeOf \app P \leadsto P': \ok 
}
{\Gamma \typeOf  (\nu a:T).P \leadsto  (\nu a:T).P' : \ok }
\quad\,\,
\inference
	{
	\Gamma \typeOf \app P \leadsto P : \ok 	} 
	{ \Gamma \typeOf \app !P \leadsto !P': \ok}
\\[2ex]
\ninference{ci-in}
	{\Gamma(a)  \sim \inp \cdot (T_1 \cdot \ldots \cdot T_n) \\
	\Gamma, a_1:T_1, \ldots ,a_n:T_n \typeOf \app P \leadsto P': \ok  	} 
	{ \Gamma \typeOf \app a(a_1:T_1, \ldots, a_n:T_n).P \leadsto \HI{$(a:\Gamma(a)  \Rightarrow \inp \cdot (T_1\cdot \ldots \cdot T_n))$}(a_1:T_1, \ldots, a_n:T_n).P': \ok}
\\[2ex]
\ninference{ci-out}
	{\Gamma(a)  \sim \out \cdot (\Gamma(a_1)\cdot \ldots \cdot \Gamma(a_n)) \\
	\Gamma \typeOf \app P \leadsto P': \ok  	} 
	{ \Gamma \typeOf \app \overline{a} \langle a_1, \ldots a_n\rangle.P \leadsto \HI{$\overline{(a:\Gamma(a)  \Rightarrow \out \cdot (\Gamma(a_1)\cdot \ldots \cdot \Gamma(a_n)))}$} \langle a_1, \ldots, a_n\rangle.P' : \ok}
\\[2ex]
\ninference{ci-rout}
	{\Gamma(a)  \sim \out \cdot (\Gamma(a_1)\cdot \ldots \cdot \Gamma(a_n)) \\
	\Gamma \typeOf \app P \leadsto P': \ok  	} 
	{ \Gamma \typeOf \app \stackrel{\rule{0.2cm}{0.4pt}_{{\mathcal{R}}}\hspace{-0.2cm}}{{a}}\langle a_1, \ldots a_n\rangle.P \leadsto \HI{$\overline{(a:\Gamma(a)  \Rightarrow \out \cdot (\Gamma(a_1)^{-1}\cdot \ldots \cdot\Gamma(a_n)^{-1}}))$} \langle a_1, \ldots , a_n\rangle.P' : \ok}
\end{gather*}
\caption{Cast Insertion for Gradually Typed Capabilities}
\label{fig:ci}
\end{figure}

\subsection{Operational Semantics for Channel Casts}

Below we show the syntax of $\pi_{c}^{\dyn}$, our $\pi$-calculus with cast channels. The most relevant parts are highlighted. 
\begin{syntax}
  &&&\\
  \text{\sf Types} & T,S & ::= &  I \cdot (T_1, \ldots, T_n) \mid \dyn\\
  \text{\sf Capabilities} & I & ::= &  \inp \mid \out \\
  \text{\sf Cast Channels} & c & ::= & a \mid \HI{$(c : T \Rightarrow T)$} \\
   \text{\sf Processes} & P,Q & ::= & \nilProcess ~\mid~ (\nu a:T).P ~\mid~ \HI{$c$}(a_1:T_1, \ldots a_n:T_n).P ~\mid~ \HI{$\overline{c} \langle c_1, \ldots ,c_n\rangle.P$} \\
   &&& ~\mid~ P + P ~\mid~ P \mid P ~\mid~ !P ~\mid~ \textit{typeError} \\
   \text{\sf Type Environment} & \Gamma & ::= & \emptyset \mid \Gamma, a:T
   \end{syntax}

Casts can be wrapped around channels, and, as a matter of fact, around sequences of cast channels. We can therefore have a channel being the subject of several consecutive casts. 
Cast channels can be used as the subject of input and output operations. Furthermore, channels can be communicated even if they have casts around. 

Fig \ref{fig:reduction} contains the operational semantics of $\pi_{c}^{\dyn}$. This semantics extends that of the $\pi$-calculus by adding reduction rules to handle casts on channels. Fig \ref{fig:reduction} makes use of the structural congruence $\equiv$, which is standard and has been omitted. In the following, we comment on the relevant parts of the operational semantics. 

\begin{figure}[tbp]
\small
\textsf{Reduction Semantics}  \hfill \fbox{$P \step P$}
\begin{gather*}
\inference
	{P_1\step P_1'}
	{P_1\mid P_2 \step P_1'\mid P_2}
	\qquad
\inference
	{P\step P'}
	{(\nu a:T).P \step  (\nu a:T).P'}
\\[2ex]
\inference
	{P \equiv P' & P' \step Q' & Q' \equiv Q}
	{P \step Q}
\\[2ex]
\ninference{comm}{}
	{
	a(a_1:T_1, \ldots, a_n:T_n).P ~\mid~ \overline{a} \langle c_1, \ldots, c_n\rangle.Q  ~\step~  P\{c_1/a_1, \ldots, c_n/a_n\} ~\mid~ Q
	}
\\[2ex]
\ninference{c-solve}
{\textit{either $c_1$ or $c_2$ is not a channel (i.e. it has a cast)}\\
\key{ch}(c_1) = \key{ch}(c_2) \\\\
\overline{c_2} \langle c_1', \ldots, c_n'\rangle.Q ~\step\subCast~  Q' \\\\
 c_1(a_1:S_1, \ldots, a_n:S_n).P \mid~ Q' ~\step\subCast~  R }
{	c_1(a_1:S_1, \ldots, a_n:S_n).P  \mid~ \overline{c_2} \langle c_1', \ldots, c_n'\rangle.Q  ~\step~  R}
\end{gather*}
\textsf{Cast Reduction}  \hfill \fbox{$P \step\subCast P$}
\begin{gather*}
\ninference{c-out-base}{}
{\overline{a} \langle c_1, \ldots, c_n\rangle.P ~\step\subCast~  \overline{a} \langle c_1, \ldots, c_n\rangle.P}
\\[2ex]
\ninference{c-out-succeed}{\overline{c} \langle (c_1 : S_1 \Rightarrow T_1), \ldots, (c_n : S_n \Rightarrow T_n)\rangle.P ~\step\subCast~ R}
{	\overline{(c : \out \cdot (T_1\cdot \ldots\cdot T_n) \Rightarrow  \out \cdot (S_1\cdot \ldots\cdot S_n))} \langle c_1, \ldots, c_n\rangle.P  ~\step\subCast~ R}
\\[2ex]
\ninference{c-out-fail}{}
{	\overline{(c : \inp \cdot (T_1\cdot \ldots\cdot T_n) \Rightarrow  \out \cdot (S_1\cdot \ldots\cdot S_n))} \langle c_1, \ldots, c_n\rangle.P  ~\step\subCast ~\typeError}
\\[2ex]
\ninference{c-out-expand}
{\overline{(c : \out \cdot (\dyn\cdot \ldots\cdot\dyn) \Rightarrow \out \cdot (S_1\cdot \ldots\cdot S_n))} \langle c_1, \ldots, c_n\rangle.P \step\subCast R}
{	
	\overline{(c : \dyn \Rightarrow \out \cdot (S_1\cdot \ldots\cdot S_n))} \langle c_1, \ldots, c_n\rangle.P  ~\step\subCast~ R}
\\[2ex]
\ninference{c-in-base}{}
{a(a_1:S_1, \ldots, a_n:S_n).P \mid~ \overline{a} \langle c_1, \ldots, c_n\rangle.Q ~\step\subCast~ a(a_1:S_1, \ldots, a_n:S_n).P \mid~ \overline{a} \langle c_1, \ldots, c_n\rangle.Q}
\\[2ex]
\ninference{c-in-succeed}
{c(a_1:T_1, \ldots, a_n:T_n).P  \mid~ \overline{a} \langle (c_1:S_1\Rightarrow T_1), \ldots, (c_n:S_n\Rightarrow T_n)\rangle.Q ~\step\subCast~ R}
{(c : \inp \cdot (T_1\cdot \ldots\cdot T_n) \Rightarrow  \inp \cdot (S_1\cdot \ldots\cdot S_n))(a_1:S_1, \ldots, a_n:S_n).P  \mid~ \overline{a} \langle c_1, \ldots, c_n\rangle.Q  ~\step\subCast~ R}
\\[2ex]
\ninference{c-in-fail}{}
{(c : \out \cdot (T_1\cdot \ldots\cdot  T_n) \Rightarrow  \inp \cdot (S_1\cdot \ldots\cdot S_n))(a_1:S_1, \ldots, a_n:S_n).P  \mid~ \overline{a} \langle c_1, \ldots, c_n\rangle.Q  ~\step\subCast~ \typeError}
\\[2ex]
\ninference{c-in-expand}
{(c :\inp \cdot (\dyn \cdot \ldots\cdot \dyn) \Rightarrow  \inp \cdot (S_1\cdot \ldots\cdot S_n))(a_1:S_1, \ldots, a_n:S_n).P  \mid~ \overline{a} \langle c_1, \ldots, c_n\rangle.Q ~\step\subCast~  R}
{(c :\dyn \Rightarrow  \inp \cdot (S_1\cdot \ldots\cdot S_n))(a_1:S_1, \ldots, a_n:S_n).P  \mid~ \overline{a} \langle c_1, \ldots, c_n\rangle.Q  ~\step\subCast~ R}
\end{gather*}
\caption{Operational Semantics for the $\pi$-calculus with Cast Channels. Notice the absence of a reduction rule for $P + P'$: this is because, in this preliminary work, we let it reduce to either $P$ or $P'$ as needed (cf. Remark \ref{remark}).}
\label{fig:reduction}
\end{figure}

\textsc{(comm)} is the standard reduction rule for communication. We point out that it can fire only so long as the input and the output have a bare channel (no casts) as subject. Therefore, this rule cannot fire unless casts have been previously resolved by other reduction rules. We also point out that the rule allows for cast channels to be sent, and they are substituted in the body of the receiver with the usual parameter passing mechanism. 

\textsc{(c-solve)} applies when the subjects of the input and the output happen not to be ready because wrapped by some casts. In that case, we first need to check that casts are successful. 
As we shall see in the context of the other reduction rules, handling a successful cast may mean that nested casts must be distributed to the arguments of outputs. Small-step semantics is not ideal because in one step we may choose a communicating partner, and the next step we may choose a completely different output partner. It is incorrect, however, to partially distribute some casts to one process and some others to another. Therefore, \textsc{(c-solve)} makes use of big-step semantics to commit to two communicating partners and resolve their casts in one big step. 
The \key{ch} function retrieves the channel name that is used at the bottom of the cast. We use \key{ch} to select two processes, one in input and one in output, that communicate through the same channel. Afterwards, the big-step reduction relation $\step^c$ solves the casts of the output and the casts of the input. These are handled, in specific cases, by the reduction rules that we describe below. 

%

\textsc{(c-out-succeed)} applies when the subject of an output is wrapped in a cast. When the channel that is buried inside the cast does have output capabilities the cast is successful. Accordingly, we simply remove the cast. However, we cannot forget that the type of the channel had arguments, which too are supposed to match the requested types. Therefore, we distribute those casts to the  arguments of the output. It is interesting to see that the direction of the cast is contravariant. 
It is so to ensure type preservation, as discussed in Section \ref{properties}. 

\textsc{(c-out-fail)}, tool, applies when the subject of an output is wrapped in a cast. This rule detects a cast failure, that is, the channel that is buried inside the cast is not prescribed for outputs. In this case we trigger a run-time type error. 

\textsc{(c-out-expand)}, similarly, applies when the subject of an output is wrapped in a cast. Since channels may be simply declared to be of dynamic type $\dyn$, we may have casts from $\dyn$ to $\out \cdot (T_1, \ldots, T_n)$. In this case, we take a step to treat $\dyn$ as $\out \cdot (\dyn, \ldots, \dyn)$, so that \textsc{(c-out-succeed)} can take over, succeed for now, but handle the rest of casts. 
(Casts from $\out \cdot (T_1, \ldots, T_n)$ to $\dyn$ do not happen in the context of this rule because the top level operation is an output and the cast insertion is guaranteed to produce a cast ending in \out.)

%
%

Rules \textsc{(c-in-*)} play the same role as rules \textsc{(c-out-*)} but they work with inputs. A key difference is that when casts are resolved their nested casts are transferred to the arguments of the output partner.

\section{Example}
\label{example}

In this section, we show how our formulation applies to the example of the revenue agency of Section \ref{background}.

\paragraph{Gradual Type Checking}

The revenue agency $A$ and the tax payer client $C$ share the initial channel $r$. 
$C$ is type checked under the assignment that $r$ must be used input-only, and used to receive a channel whose capabilities are not known at compile-time, that is, it is of dynamic type. 
Therefore, we type check $C$ with $\Gamma = r:\inp \cdot \dyn$. Let us recall $C$. ($T$ is the type of the channel representing  \$100, irrelevant here).

\begin{equation}
r(b:\dyn). ~ (\overline{b}\langle \$100\rangle ~+~ b(sum : T))\\
\end{equation}

The first rule that applies is \textsc{(t-in)} for the top level process $r(b:\dyn).\textit{(the rest of the client)}$. 
This rule applies because we are called to check $\inp \cdot \dyn \sim \inp \cdot \dyn$, which certainly holds because $\sim$ is reflexive. Afterwards, $\Gamma$ is augmented with the information about $b$ and becomes $\Gamma = r:\inp \cdot \dyn, b:\dyn$. The first branch of the choice operator $\overline{b}\langle \$100\rangle$ type checks successfully because the type $\out \cdot T$ is consistent with the type of $b$ in $\Gamma$. This check is $\dyn \sim  \out \cdot T$, which holds because $\dyn$ is consistent with every type. For analogous reasons, the second branch $b(sum : T)$ type checks successfully because we ultimately need to check $\dyn \sim  \inp \cdot T$, which holds. 

The agency is type checked under the assignment that $r$ must be used output-only for sending a channel of dynamic type, that is, $\Gamma = r:\out \cdot \dyn$.
\begin{flalign}
A  \equiv\app &\nu(x:\out \cdot T).\stackrel{\rule{0.2cm}{0.4pt}_{{\mathcal{R}}}\hspace{-0.2cm}}{{r}}\langle x\rangle.\overline{x}\langle \$100\rangle \\\nonumber
&+\\\nonumber
&\nu(x:\inp \cdot T).\stackrel{\rule{0.2cm}{0.4pt}_{{\mathcal{R}}}\hspace{-0.2cm}}{{r}}\langle x\rangle.x(sum)
\end{flalign}

Let us consider the first branch of the choice operator. \textsc{(t-res)} introduces $x$, making $\Gamma = r:\out \cdot \dyn, x:\out \cdot T$.
To type check the reverse output, we need to consider its type $\out \cdot \HI{$(\out \cdot T)$}$, where the highlighted part is grabbed from the information of $x$ in $\Gamma$. This type must be compared with the type of $r$ which is $r:\out \cdot \dyn$. The check is $\out \cdot \dyn \sim \out \cdot (\out \cdot T)$, which holds. Finally, $\overline{x}\langle \$100\rangle$ is type checked successfully because we end up checking $\out \cdot T \sim \out \cdot T$, which holds by reflexivity. 
The second branch of the choice operator follows similar lines. 

\paragraph{Cast Insertion}

We start with the client. For the top level process $r(b:\dyn).\textit{(the rest of the client)}$, the relation $\inp \cdot \dyn \sim \inp \cdot \dyn$ induces the cast $(r:\inp \cdot \dyn  \Rightarrow \inp \cdot \dyn) (b:\dyn).\textit{(the rest of the client)}$. However, this is a trivial cast and we omit it, as implementations in fact do skip trivial casts.  
The first branch of the choice operator $\overline{b}\langle \$100\rangle$ prescribes checking $\dyn \sim  \out \cdot T$, which induces the cast $\overline{(b:\dyn \Rightarrow \out \cdot T)}\langle \$100\rangle$. The second branch of the choice operator prescribes checking $\dyn \sim  \inp \cdot T$, which induces the cast $(b:\dyn \Rightarrow \inp \cdot T)(sum))$. The result of the compilation for the client is: 
\begin{gather*}
C' \equiv \app r(b:\dyn). ~ (\overline{(b:\dyn \Rightarrow \out \cdot T)}\langle \$100\rangle ~+~ (b:\dyn \Rightarrow \inp \cdot T)(sum))
\end{gather*}

Let us consider the compilation for the first branch of the choice operator of the agency. Going through a restriction $\nu$ does not generate new casts. Afterwards, the use of the reverse output does generate casts. The checking that takes place is $\out \cdot \dyn \sim \out \cdot (\out \cdot T)$. Therefore, we apply the cast insertion rule for reverse output and we have $(\overline{r:\out \cdot \dyn \Rightarrow \out \cdot (\HI{$\inp$} \cdot T)})\langle x\rangle$. Notice the highlighted input capability. This type comes from $\Gamma(a)^{-1}$ when applied to $(\out \cdot T)$, which is $(\inp \cdot T)$. Notice also that the reverse output became an ordinary output. The compilation for the second branch of the choice operator is similar. Ultimately, the compilation of the agency process is the following. 
\begin{flalign}
A' \equiv & ~\nu(x:\out \cdot T).(\overline{r:\out \cdot \dyn \Rightarrow \out \cdot (\inp \cdot T)})\langle x\rangle.\overline{x}\langle \$100\rangle \\\nonumber
&+\\\nonumber
&\nu(x:\inp \cdot T).(\overline{r:\out \cdot \dyn \Rightarrow \out \cdot (\out \cdot T)})\langle x\rangle.x(sum)
\end{flalign}

\noindent These casts are slightly different from those of the example in Section \ref{background}. One reduction step will distribute these casts to the arguments, reaching that form. We shall see this aspect in the next paragraph. 


\paragraph{Execution}

Let us consider the execution of the compiled agency and the compiled client in parallel, i.e., $(A' \mid C')$. 
Let us suppose that we are in the scenario in which the tax payer must send a payment\footnote{Recall that, for simplicity, we assume that the $+$ operator will select the right branch for us, cf. Remark \ref{remark}.}.
Structural congruence brings together the two processes so that ultimately the rule \textsc{(comm)} would apply: 
\[
r(b:\dyn). ~ (\overline{(b:\dyn \Rightarrow \out \cdot T)}\langle \$100\rangle ~+~ (b:\dyn \Rightarrow \inp \cdot T)(sum))
\mid 
(\overline{r:\out \cdot \dyn \Rightarrow \out \cdot (\out \cdot T)})\langle x\rangle.x(sum)
\]
However, \textsc{(comm)} cannot apply just yet because the rule works only when the subject of inputs and outputs are bare channels. Instead, rule \textsc{(c-solve)} applies. This rule has the effect of applying \textsc{(c-out-succeed)} to the output process. This step detects that the top level cast is an $\out$-to-$\out$ match, and so it removes the cast and distributes the nested casts to the argument:
\[
(\overline{r:\out \cdot \dyn \Rightarrow \out \cdot (\inp \cdot T)})\langle x\rangle.\overline{x}\langle \$100\rangle \step\subCast
\overline{r}\langle (x : \out \cdot T \Rightarrow \dyn)\rangle.x(sum) \\\nonumber
\]

\noindent Therefore, in one step we have a process that is in a form that \textsc{(comm)} handles: 
\[
r(b:\dyn). ~ (\overline{(b:\dyn \Rightarrow \out \cdot T)}\langle \$100\rangle ~+~ (b:\dyn \Rightarrow \inp \cdot T)(sum))
\mid 
\overline{r}\langle (x : \out \cdot T \Rightarrow \dyn)\rangle.x(sum) 
\]
\noindent And in one step we obtain:  
\[
\overline{(x : \out \cdot T \Rightarrow \dyn \Rightarrow \out \cdot T)}\langle \$100\rangle ~+~ (x : \inp \cdot T \Rightarrow \dyn \Rightarrow \inp \cdot T)(sum)
\mid 
x(sum) 
\]
At this point, structural congruence brings the output on $x$ and its input $x(sum)$ together. Since the output on $x$ has casts, we have that \textsc{(c-solve)} applies. This rule has the effect to first apply \textsc{(c-out-expand)} because the top level cast is from $\dyn$ to $\out \cdot T$. This application of \textsc{(c-out-expand)} expands the inner cast and turns the output into $\overline{(x : \out \cdot T \Rightarrow \out \cdot \dyn \Rightarrow \out \cdot T)}\langle \$100\rangle$. At this point, two applications of \textsc{(c-out-succeed)} push these casts to the argument. This is possible because all casts are an $\out$-to-$\out$ match, and therefore the rule applies. Ultimately, we end up with the following process (isolated by structural congruence for execution). 

\[
\overline{x}\langle \$100 : T \Rightarrow \dyn \Rightarrow T\rangle
\mid 
x(sum)
\]
Here, \textsc{(comm)} applies as usual. Notice that $\$100 : T \Rightarrow \dyn \Rightarrow T$ is not resolved immediately. It will be checked when $\$100 : T \Rightarrow \dyn \Rightarrow T$ is going to be used, \emph{and if} it is going to be used. This is in line with the dynamic typing style. Such event does not occur in our example, but we could imagine that happening if $x(sum)$ had a continuation process. 

\section{Properties (Not Addressed in This Preliminary Work)}
\label{properties}

In this paper, we do not claim any theoretical results about our formalisms, hence the ``\emph{Towards}'' part of the title. 
In future we would like to prove some key properties of our formulations. 
The properties that we plan on attacking can be divided into two kinds. 
The first kind concerns the typical properties at play for typed calculi: the \emph{progress theorem} and the \emph{type preservation theorem}. The second kind of properties are specific to the domain of gradual typing. 
In this section we discuss some of these properties. 

\paragraph{Progress}
For the progress theorem, we want to prove that if a process $P$ is well-typed then either $P = typeError$, 
or $P$ is stuck in a situation in which no communication can take place (as in $a(x).P \mid b(x).P$), or $P \step P'$, for some $P'$. To prove this property we need to check that all behavior is covered. 
We make an example of the reasoning that applies in our context. 

The only additions to ordinary $\pi$-calculus is that we can have casts in 3 places: 

\begin{itemize}
\item a cast for a channel being sent, as in $\overline{a}\langle c : \dyn \Rightarrow \out \rangle.P$. If this output does not have an input partner it would be correctly stuck in our calculus. If it does have a communicating partner then the ordinary communication rule of the $\pi$-calculus takes place with no restriction, therefore it progresses.
%
\item a cast for a channel that is subject of an output, as in $\overline{a : \textit{list of casts}}\langle c \rangle.P$. In this case, rules \textsc{(c-out-succeeds)}, \textsc{(c-out-fail)}, and \textsc{(c-out-expand)} of Fig. 4 apply in all possible cases, depending on the casts. Indeed, a cast can only be $\out$-to-$\out$, handled by \textsc{(c-out-succeeds)}, $\inp$-to-$\out$, handled by \textsc{(c-out-fail)}, or  $\dyn$-to-$\out$, handled by \textsc{(c-out-expand)}, with the latter leading directly to a form handled by \textsc{(c-out-succeeds)}. 
Casts to $\inp$ or $\dyn$ cannot happen because the cast insertion always places a cast to $\out$ in the context of an output. 
%
\item a cast for a channel that is subject of an input, as in $(a : \textit{list of casts})(b:T).P$. Here casts are resolved in essentially the same way as in the previous case, though dually. 
\end{itemize}

\paragraph{Type Preservation}

We would like to prove that if a process $P$ is well-typed and $P \step P'$ then $P'$ is well-typed, as well. 
This would require a case analysis on the steps that $P$ can take. Especially, the new reduction rules that handle casts will be the relevant part of this proof. We make an example of the reasoning that applies in our context, just for one reduction rule: \textsc{(c-out-succeed)}.
Let us consider the step that \textsc{(c-out-succeed)} provides. 
Suppose $a$ is an output channel that sends one channel of dynamic type, and $b$ is an input channel. 
Therefore, we have $\Gamma = a : \out \cdot \dyn, b : \inp$.   
Suppose we have the process $(\overline{a : \out \cdot \dyn \Rightarrow \out \cdot \inp})\langle b \rangle$.
This is a well-typed process because the cast channel used in output is ultimately of type $\out \cdot \inp$ (the target of the cast), and $b$ is indeed of type $\inp$. Therefore, types all match. 
After the step of \textsc{(c-out-succeed)} we have the cast pushed into the argument: $\overline{a}\langle b : \inp \Rightarrow \dyn\rangle$. We have to check that this process is well-typed. It is, indeed: $a$ is a channel of type $\out \cdot \dyn$ and it is used in output. We have to check that the sent channel is of type $\dyn$. This is the case because the sent argument is $b : \inp \Rightarrow \dyn$, which ultimately is of type $\dyn$ because of the cast to $\dyn$. 
%
%

\paragraph{Gradual Typing Properties}
In future, we would like to address the correctness criteria of gradual typing as summarized and delineated in \cite{Siek2015snapl}.
In this paragraph, we discuss only some relevant ones. 

 A property that we want to prove is that our type system is a conservative extension of that of Pierce and Sangiorgi (when $\either$ is not used). We expect this property to hold because the type consistency $\sim$ is designed to cover the same cases of Pierce's and Sangiorgi's type system, and \emph{additionally} some more. 
We have to be careful, however, that the additional processes that we can type check are not incorrectly deemed well-typed. 
It would be incorrect for example to type check $a(b:\out).b(z)$. This would not happen in our calculus despite $\sim$ being a liberal relation because it still mandates $\inp\not\sim\out$, and thus would reject that process. 

Similarly, also the operational semantics should be a conservative extension of that of the $\pi$-calculus. That is, processes that do no contain casts at all should be executed as they were in the ordinary $\pi$-calculus. 
We expect this property to hold because when casts are not around then the plain reduction rules of the $\pi$-calculus, and solely those, apply.

Another property that we would like to prove is that our gradualized type system is monotonic w.r.t. the amount of type annotations that are turned into the dynamic type.  
This is a fundamental property in gradual typing: To make an example, if we were in the $\lambda$-calculus, this property would entail that we can write a generalized identity function $(\lambda x : \dyn . x)$ and it \emph{must} be well-typed (as we expect), because $(\lambda x : \Int . x)$ is well-typed, and the former simply turns a type into the dynamic type. 
We expect this property to hold in our type system because the type consistency $\sim$ is indeed designed to allow type checking to succeed when the dynamic type is met. 


The converse monotonic theorem, in which we turn dynamic types into specific types and expect to preserve typeability, does not, and should not, hold: it is not guaranteed that if a programmer inserts new type annotations then the process is well-typed. Indeed, the programmer might insert a wrong type.

We would like to address the blame theorem \cite{AhmedFSW11,WadlerF09,Tobin-Hochstadt:2006}. This property states that run-time cast errors can happen only in the regions of code that are dynamically typed. To address this property we will need to modify our calculus to keep, at run-time, the information of a label on casts. This label pinpoints the origin of the cast in the source code. Furthermore, our operational semantics must be equipped with a suitable mechanism for blame tracking.

We also would like to address the gradual guarantee. This property states that adding types to a gradually typed process has the effect that the new version of the process can only i) end up ill-typed at compile-time, ii) behave the same way as the previous version, or iii) end up in a cast error at run-time (if the new type inserted was wrong but this could not be detected at compile-time). 
This is an important property that provides programmers with the guarantee that adding types would not corrupt unpredictably the execution of processes. As a consequence, gradual typing could be used to evolve a process into a more and more statically typed version over time without disrupting the workflow of the programmer. We expect this property to be challenging, as it has been proven challenging in several contexts \cite{Siek2015snapl,IgarashiSI17,Cimini:2017}. 



\section{Related Work and (More) Future Work}
\label{related}

The work that is most related to that of this paper is the typed $\pi$-calculus of Pierce and Sangiorgi \cite{tp}. 
We have delineated the major differences with our calculus in Section \ref{background} and Section \ref{gt}. 
An additional difference is that the typed $\pi$-calculus makes use of recursive types to type check processes such as $\overline{a}\langle a \rangle$. We have omitted recursive types but our type system can still type check those cases by giving the channel $a$ the type $\dyn$. However, adding recursive types is part of our future work so that we would be able to describe more precisely cases of the like.  
Recursive types have been addressed previously in gradual typing implementations \cite{Hejlsberg:2012aa,track,andre,KuhlenschmidtAS19}, and in semantics formulations \cite{rec,NewLA19}. We plan on building on this body of work.  

We have removed the type $\either$ altogether, which means that our channels can be used input-only or output-only. 
The feature that $\either$ offers cannot be recovered completely with $\dyn$ because once the input or output capability of a channel has been established at run-time it cannot be used for the opposite operation. 
We therefore plan to extend our formulations to have both $\either$ and $\dyn$. 

There has been considerable work in gradual typing both from industry  \cite{Bierman:2014aa,Hejlsberg:2012aa,Verlaguet:2013aa,Chaudhuri:2014aa} and academia \cite{Siek:2006bh,Tobin-Hochstadt:2006,ToroGT18,LehmannT17,SchwerterGT16,AhmedFSW11}. 
%
%
The challenge in capturing a calculus such as the $\pi$-calculus is that several output processes may compete for a single input. We, however, must stick to one such communicating process after initiating a communication, or otherwise we would transfer some casts to a process and some others to another. Because of this, we have used big-step semantics to commit to a communicating process throughout solving all casts. To our knowledge, this is an aspect that has not been previously addressed in gradual typing.


Gradual typing has been applied to process calculi previously, and in particular to gradual session types \cite{Igarashi:2017,thiemannST}. The major differences between gradual session types and the work of this paper are: 

\begin{itemize}
\item Types in \cite{Igarashi:2017,thiemannST} are set to enforce session fidelity: every send is matched with a receive, every select is matched with an offer, and so on, and deadlock freedom is ensured. The type system we focus on is derived from that of Pierce and Sangiorgi, which does not strictly structure the communication taking place. It simply keeps track of the capabilities of channels. Therefore, our type system is more permissive, and describes more computations, but is also less safe. 
In our calculus we can describe a race condition with one input and several output partners, for example. Similarly, we can type check a single process in input hanging on forever without outputs. These scenarios are rejected in gradual session types but are common in Pierce's and Sangiorgi's calculus because, as exemplified in Section \ref{background}, the type system is frequently used to type check only a part of the whole process. 

\item Both \cite{Igarashi:2017} and \cite{thiemannST} only focus on dyadic session types, that is, 2 dually communicating processes. Our calculus allows for an unrestricted number of processes in parallel. This raised the challenge described above on committing to a communicating process, which we have solved with a big-step semantics.
\end{itemize}

There has been some work in automating the shift to gradual typing \cite{Cimini:2016,Cimini:2017,Garcia:2016}. We have found out that because of the particular challenge with committing to a communicating process these frameworks could not be applied. That is to say that our formulations could not be generated automatically with current automated techniques. 

\section{Conclusions}
\label{conclusions}

We have presented our preliminary work towards a calculus of gradual capabilities for the $\pi$-calculus. 
We have formulated a gradual type system, a cast insertion procedure, and a $\pi$-calculus that can handle casts on channels. 
We do not claim any theoretical results about our formalisms at this stage. We have shown how our formalisms apply to an example that requires the discovery of channel capabilities at run-time. We also have discussed our future plans, especially w.r.t. the key properties that a gradually typed calculus such as ours should afford. 

\bibliographystyle{splncs04}
\bibliography{local.bib} 

\end{document}

%% file: definitions.tex
\newcommand*\colourcheck[1]{%
  \expandafter\newcommand\csname #1check\endcsname{\textcolor{#1}{\text{\ding{52}}}}%
}
\colourcheck{blue}
\colourcheck{green}
\colourcheck{red}

\newcolumntype{Y}{>{\raggedleft\arraybackslash}X}

\tcbset{tab1/.style={fonttitle=\bfseries\large,fontupper=\normalsize\sffamily,
colback=yellow!10!white,colframe=red!75!black,colbacktitle=red!40!white,
coltitle=black,center title,freelance,frame code={
\foreach \n in {north east,north west,south east,south west}
{\path [fill=red!75!black] (interior.\n) circle (3mm); };},}}

\tcbset{tab2/.style={enhanced,fonttitle=\bfseries,fontupper=\small\sffamily,
colback=yellow!10!white,colframe=red!50!black,colbacktitle=red!40!white,
coltitle=black,center title}}

\usepackage{relsize}

\newcommand{\inp}{\key{i}}
\newcommand{\out}{\key{o}}
\newcommand{\either}{\key{either}}
\newcommand{\ok}{\key{ok}}
\newcommand{\nilProcess}{\textbf{0}}
\newcommand{\typeError}{typeError}

\newcommand{\subCast}{_{\textsf{c}}}

\newcommand{\ninference}[3]{\inferrule[(#1)]{#2}{#3}}

\newcommand{\ba}{\begin{array}}
\newcommand{\ea}{\end{array}}

\newenvironment{syntax}{\[\ba{l@{\;\;}lcl}}{\ea\]}

\definecolor{ShadowColor}{RGB}{30,150,190}

\makeatletter
\newcommand\Cshadowbox{\VerbBox\@Cshadowbox}
\def\@Cshadowbox#1{%
  \setbox\@fancybox\hbox{\fbox{#1}}%
  \leavevmode\vbox{%
    \offinterlineskip
    \dimen@=\shadowsize
    \advance\dimen@ .5\fboxrule
    \hbox{\copy\@fancybox\kern.5\fboxrule\lower\shadowsize\hbox{%
      \color{ShadowColor}\vrule \@height\ht\@fancybox \@depth\dp\@fancybox \@width\dimen@}}%
    \vskip\dimexpr-\dimen@+0.5\fboxrule\relax
    \moveright\shadowsize\vbox{%
      \color{ShadowColor}\hrule \@width\wd\@fancybox \@height\dimen@}}}
\makeatother

\newcommand{\typeOf}{\vdash}
\newcommand{\step}{\longrightarrow}

\definecolor{lightblue}{rgb}{0.25,0.25,1}

\definecolor{lightgray}{gray}{0.9}
\definecolor{darkergrey}{rgb}{0.75, 0.75, 0.75}
\newcommand{\HI}[1]{\colorbox{darkergrey}{#1}}

\newcommand{\key}[1]{\ensuremath{\mathtt{#1}}}

\newcommand{\Int}{\key{Int}}

\newcommand{\app}{\;}

\definecolor{lightgray}{gray}{0.9}

\newcommand\dyn{\star}

\newcommand{\Ldl}{\mathcal{L}}

\newcommand{\emptyLdl}[1]{\Ldl^{\epsilon}}